\documentclass[%
 reprint,
 superscriptaddress,
 amsmath,amssymb,
 aps,
 floatfix,
]{revtex4-2}
\usepackage{siunitx}
\usepackage{graphicx}
\usepackage{dcolumn}
\newcommand{\RNum}[1]{\uppercase\expandafter{\romannumeral #1\relax}}
\usepackage{wrapfig}
\usepackage{float}
\usepackage{bm}
\usepackage[dvipsnames]{xcolor}
\usepackage{hyperref}

\begin{document}
\preprint{APS/123-QED}
\title{First principles feasibility assessment of a topological insulator at the InAs/GaSb interface}
\author{Shuyang Yang}
 \affiliation{Department of Materials Science and Engineering, Carnegie Mellon University, Pittsburgh, PA 15213, USA}
 \author{Derek Dardzinski}
 \affiliation{Department of Materials Science and Engineering, Carnegie Mellon University, Pittsburgh, PA 15213, USA}
\author{Andrea Hwang}
 \affiliation{Department of Materials Science and Engineering, Carnegie Mellon University, Pittsburgh, PA 15213, USA}
 \author{Dmitry I. Pikulin}
 \affiliation{Microsoft Quantum, Redmond, WA 98052, USA}
\affiliation{Microsoft Quantum, Microsoft Station Q, University of California, Santa Barbara, California 93106-6105, USA}
 \author{Georg W. Winkler}
\affiliation{Microsoft Quantum, Microsoft Station Q, University of California, Santa Barbara, California 93106-6105, USA}
\author{Noa Marom}
\email{nmarom@andrew.cmu.edu}
 \affiliation{Department of Materials Science and Engineering, Carnegie Mellon University, Pittsburgh, PA 15213, USA}
 \affiliation{Department of Chemistry, Carnegie Mellon University, Pittsburgh, PA 15213, USA}
\affiliation{Department of Physics, Carnegie Mellon University, Pittsburgh, PA 15213, USA}
\date{\today}
\newcommand{\sy}[1]{{\color{blue} Shuyang: #1}}
\begin{abstract}
First principles simulations are conducted to shed light on the question of whether a two-dimensional topological insulator (2DTI) phase may be obtained at the interface between InAs and GaSb. To this end, the InAs/GaSb interface is compared and contrasted with the HgTe/CdTe interface. Density functional theory (DFT) simulations of these interfaces are performed using a machine-learned Hubbard U correction [npj Comput. Mater. 6, 180 (2020)]. For the HgTe/CdTe interface our simulations show that band crossing is achieved and an inverted gap is obtained at a critical thickness of 5.1 nm of HgTe, in agreement with experiment and previous DFT calculations. In contrast, for InAs/GaSb the gap narrows with increasing thickness of InAs; however the gap does not close for interfaces with up to 50 layers (about 15 nm) of each material. When an external electric field is applied across the InAs/GaSb interface, the GaSb-derived valence band maximum is shifted up in energy with respect to the InAs-derived conduction band minimum until eventually the bands cross and an inverted gap opens. Our results show that it may be possible to reach the topological regime at the InAs/GaSb interface under the right conditions. However, it may be challenging to realize these conditions experimentally, which explains the difficulty of experimentally demonstrating an inverted gap in InAs/GaSb.
\end{abstract}
\maketitle

\section{INTRODUCTION}

Two-dimensional topological insulators (2DTIs) have attracted increasing attention in recent years owing to the emergence of helical edge states and backscattering-free edge currents relevant for applications in spintronics and quantum computing \cite{hasan2010colloquium, ando2013topological, doi:10.1021/acs.jpclett.7b00222}.
2DTIs were first proposed based on a theoretical model of graphene incorporating spin-orbit interactions \cite{PhysRevLett.95.226801}. However, the required type of spin-orbit coupling in graphene is too weak to observe the quantum spin Hall effect (QSHE) experimentally \cite{PhysRevLett.96.106802}.
Later, a proposal for 2DTI was made based on a HgTe/CdTe quantum well (QW)~\cite{bernevig2006quantum} and the signatures of the QSHE were experimentally demonstrated~\cite{konig2007quantum}.
When the thickness of the HgTe in the QWs is varied, the band structure changes from a trivial insulator to a 2DTI with an inverted gap when a critical thickness is reached \cite{bernevig2006quantum,konig2007quantum, PhysRevB.86.075316, kufner2014topological}. 2DTIs have been proposed in additional materials systems, some of which have shown promising signs \cite{yang2012spatial,li2016experimental,wu2016evidence,pauly2015subnanometre,reis2017bismuthene, zhu2019evidence, deng2018epitaxial,wang2016topological,tang2017quantum,jia2017direct, fei2017edge, wu2018observation, sajadi2018gate}.

In the present work we focus on another QW structure, InAs/GaSb.
It has been proposed that a 2DTI may be realized in InAs/GaSb QWs because the band lineup of coupled InAs/GaSb QWs could lead to the coexistence of electrons and holes at the charge neutrality point \cite{PhysRevLett.79.3034, PhysRevB.57.11915}.
The topological insulator phase would arise if the band ordering were inverted and coupling between electron and hole states opened a hybridization gap which is necessarily topological due to the orbital structure of the hybridized bands~\cite{liu2008quantum}.
Such band ordering could potentially be achieved by choosing appropriate QW thickness and by applying an external electric field ~\cite{PhysRevLett.96.106802}.
InAs/GaSb QWs are in the family of well-studied III-V compounds and have thus attracted considerable experimental interest~\cite{PhysRevB.87.235311,PhysRevB.91.245309,PhysRevLett.107.136603,pribiag2015edge,PhysRevLett.112.026602,nichele2016edge,PhysRevLett.114.096802, PhysRevB.92.081303,shojaei2018materials}.
The experiments have provided some encouraging signs of edge conductance in the material. However, a phase diagram showing a clear topological transition accompanied by edge state formation has yet to be demonstrated. Here, we use first principles simulations to investigate whether it would be possible to realize a 2DTI at the InAs/GaSb interface and under what conditions.

The HgTe/CdTe and InAs/GaSb interfaces have been studied theoretically using a variety of methods. This includes the k$\cdot$p method \cite{sengupta2013design,kufner2014topological,PhysRevB.86.235311, PhysRevB.60.5590, li2018hidden, skolasinski2018robust}, pseudopotential models \cite{PhysRevLett.105.176805, PhysRevB.68.155329}, and tight-binding \cite{PhysRevB.76.045302, PhysRevB.69.085316}. The drawback of these semi-empirical methods is that the fitting to experimental data  largely  determines  the  extent  of  their  predictive  capability. Atomistic \textit{ab initio} simulations may provide a more accurate representation of the electronic properties and their dependence on the structure of the interface. 
First principles studies based on density functional theory (DFT) have investigated the influence of thickness on the edge states of HgTe/CdTe(100) \cite{PhysRevB.89.195312, PhysRevB.90.195311}.  Using different exchange-correlation functionals and different thicknesses of CdTe, Ref. \cite{PhysRevB.89.195312} predicted a critical thickness of 4.6 nm of HgTe, whilst Ref. \cite{PhysRevB.90.195311} predicted a critical thickness 6.5 nm of HgTe. Both results are close to the experimental critical thickness of 6.3 nm \cite{bernevig2006quantum}.  

For InAs and GaSb, local and semi-local exchange-correlation functionals severely  underestimate the band gaps to the point that they reduce to  zero \cite{10.1063/1.2404663}, due to the self-interaction error (SIE).  Some DFT studies of InAs/GaSb have applied an empirical correction to the  DFT band gaps \cite{wang2014band, sun2011first}. Others have used hybrid functionals, which mitigate the effect of SIE by including a fraction of exact exchange \cite{garwood2017electronic}. An alternative approach, which has been used to obtain more accurate band gaps for InAs/GaSb is many-body perturbation theory within the $GW$ approximation, where $G$ stands for the one-particle Green's function and $W$ stand for the screened Coulomb interaction  \cite{Taghipour_Shojaee_Krishna_2018}.  Although hybrid DFT functionals and the $GW$ approximation produce significantly improved band gaps, their high computational cost limits their applicability to relatively small system sizes. Therefore, these methods have been used only for periodic heterostructures of InAs/GaSb with very few layers \cite{garwood2017electronic, Taghipour_Shojaee_Krishna_2018}.  DFT studies of large interface slab models with vacuum regions have not been conducted. All previous \textit{ab initio} studies of InAs/GaSb have not reported the band structure and band alignment at the interface and have not shown an inverted band gap. Furthermore, previous studies have not considered the effect of applying an electric field, which plays an important role in experiments, and therefore should be considered computationally.
 
 Recently, we have introduced a new method of DFT with a machine-learned Hubbard U correction, which can provide a solution for accurate and efficient simulations of InAs and GaSb \cite{yu2020machine}. Within the Dudarev formulation of DFT+U \cite{PhysRevB.57.1505} the effective Hubbard U is defined as $U_{eff}=U-J$, where $U$ represents the on-site Coulomb repulsion, and $J$ represents the exchange interaction. For a given material, the $U_{eff}$ parameters of each element are machine-learned using Bayesian optimization (BO). The BO algorithm finds the optimal $U_{eff}$ values that maximize an objective function formulated to reproduce as closely as possible the band gap and the qualitative features of the band structure obtained with a hybrid functional. 
The DFT+U(BO) method allows for negative $U_{eff}$ values. Negative $U_{eff}$ values are theoretically permissible when the exchange term, $J$, is larger than the on-site Coulomb repulsion, $U$ \cite{RevModPhys.62.113,hase2007madelung,nakamura2009first,persson2006improved, cococcioni2012lda+}. We have found that negative $U_{eff}$ values are necessary to produce band gaps for narrow-gap semiconductors, such as InAs and GaSb. Because the reference hybrid functional calculation is performed only once for the bulk material to determine the optimal $U_{eff}$ values, the computational cost of DFT+U(BO) calculations for interfaces is comparable to semi-local DFT. 

 In this work, we use the DFT+U(BO) method to study the HgTe/CdTe and InAs/GaSb interfaces. For the HgTe/CdTe interface,
 we obtain band crossing at a critical thickness of 5.1 nm of HgTe, and subsequently an inverted gap is observed. Our results are in agreement with experiment and previous DFT studies, thus validating the DFT+U(BO) method. For the InAs/GaSb interface, we find that increasing the thickness of InAs leads to gap narrowing. However, band crossing is not obtained up to the largest number of layers calculated here. When an external electric field is applied across the InAs/GaSb interface, the GaSb-derived valence band maximum is shifted up in energy compared to the InAs-derived conduction band minimum. Band crossing is achieved at a critical field, followed by an inverted gap which widens and shifts higher above the Fermi level as the field is increased. Our results indicate that it may be possible to reach the topological regime in InAs/GaSb QWs. However, doing so would require a combination of careful interface engineering, a considerable electric field across the interface, and gating to tune the position of the Fermi level. This explains the difficulty of experimentally demonstrating an inverted gap in InAs/GaSb.

\section{METHODS}
\subsection{Computational details}

DFT calculations were performed using the Vienna \textit{ab initio} simulation package (VASP) \cite{PhysRevB.47.558} with the projector augmented wave method (PAW) \cite{PhysRevB.50.17953,PhysRevB.59.1758}. The generalized gradient approximation (GGA) of Perdew, Burke, and Ernzerhof (PBE) \cite{PhysRevLett.77.3865,PhysRevLett.78.1396} was used with a Hubbard U correction \cite{PhysRevB.57.1505} determined by Bayesian optimization \cite{yu2020machine}, as detailed below. Spin-orbit coupling (SOC) \cite{PhysRevB.93.224425} was included throughout and the energy cutoff was set to 400 eV. For bulk band structure calculations a 8$\times$8$\times$8 k-point grid was used to sample the Brillouin zone. For interface calculations a 8$\times$8$\times$1 k-point grid was used to sample the interface Brillouin zone and dipole corrections \cite{PhysRevB.46.16067} were included. Bulk unfolding \cite{yang2020electronic} was applied to project interface band structures onto the primitive cell, as described in the SI.  

\subsection{Performance of PBE+U(BO)}

\begin{figure}[H]
\includegraphics[scale = 0.07]{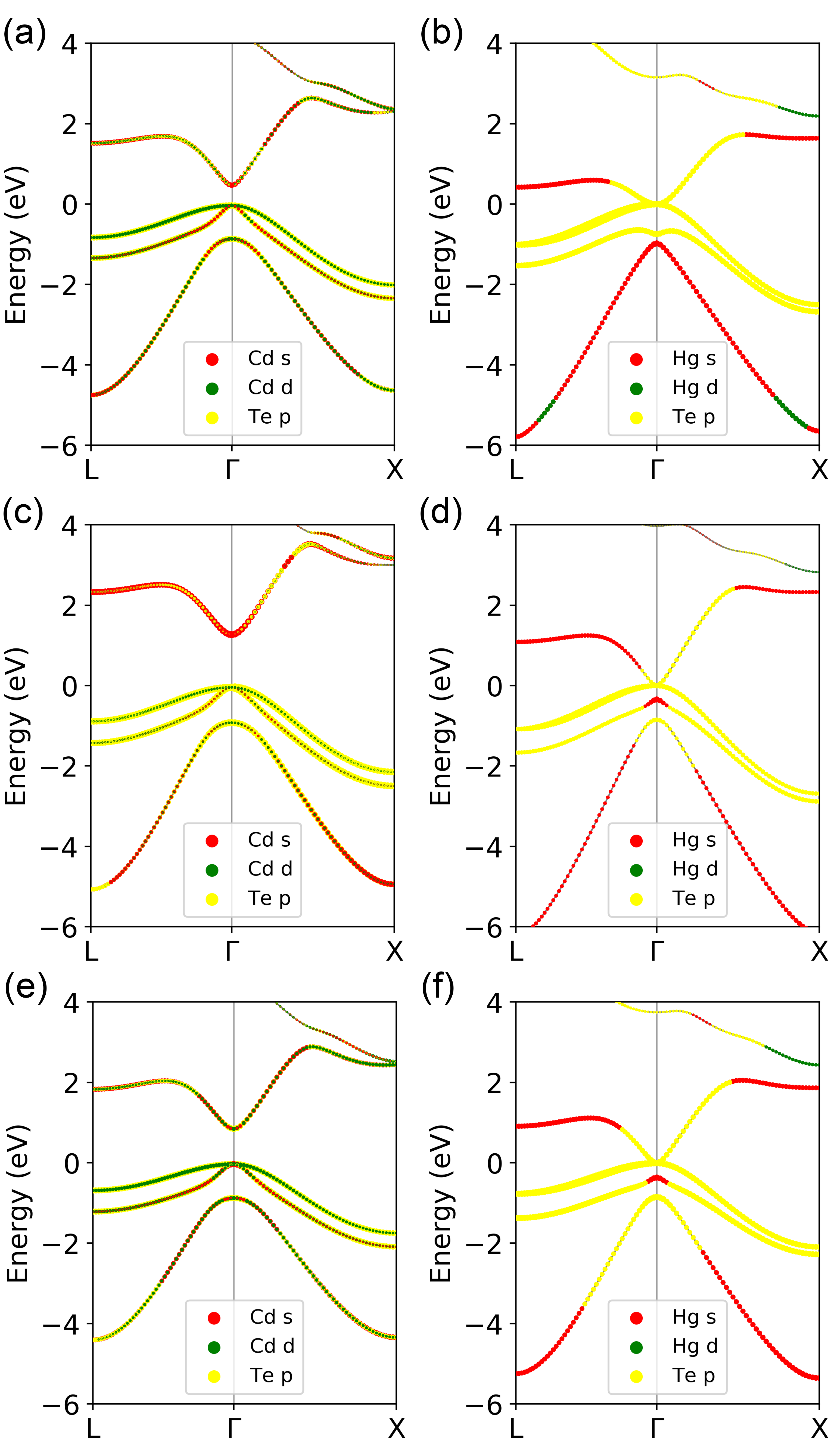}
\caption{\label{fig:hgte_hse} Performance of different DFT functionals for CdTe and HgTe: PBE band structures of (a) CdTe and (b) HgTe; HSE  band structures of (c) CdTe and (d) HgTe; PBE+U(BO) band structures of (e) CdTe and (f) HgTe; The contributions of the Cd/Hg $s$, Cd/Hg $d$, and Te $p$ states are indicated by the red, green, and yellow dots, respectively. $A$ is the point along $X-\Gamma$ in the bulk's Brillouin zone with the coordinates (0.1, 0.1, 0). $A-\Gamma-A$ is mapped to $\overline{A}-\overline{\Gamma}-\overline{A}$ in the (001) direction.}
\end{figure}

The PBE functional fails to provide an adequate description of the band structures of the materials studied here. The cases of InAs and GaSb have been discussed in detail in \cite{yu2020machine}. For CdTe, Fig. \ref{fig:hgte_hse}a shows that PBE severely underestimates the bad gap compared to the experimental value of 1.60 eV \cite{broser1982physics}. This is because the Cd 4$d$ states, which contribute significantly to the top of the valence band, are pushed up in energy due to the SIE \cite{wu2015lda+}. For HgTe, Fig. \ref{fig:hgte_hse}b shows that PBE produces an incorrect band shape and band ordering at the $\Gamma$ point with the Hg $s$ orbitals and Te $p$ orbitals inverted around 1 eV below the Fermi level \cite{PhysRevB.90.195311}. These issues are rectified by the  Heyd-Scuseria-Ernzerhof (HSE) \cite{doi:10.1063/1.2204597,Heyd2003} hybrid functional, as shown in Fig. \ref{fig:hgte_hse}c for CdTe and Fig. \ref{fig:hgte_hse}d for HgTe. However, the computational cost of HSE is too high for simulations of large interface models.

To achieve a balance between accuracy and efficiency, a Hubbard $U$ correction was applied to the $p$ orbitals of In, As, Ga and Sb and the $d$ orbitals of Hg and Cd within the Dudarev approach \cite{PhysRevB.57.1505}. For each orbital, the optimal value of U$_{eff}$ was machine learned  by Bayesian optimization \cite{yu2020machine}. The objective function was formulated to reproduce as closely as possible the band structure produced by HSE:
\begin{equation}
{f(\vec{U})}=-{\alpha_1}(\text{E}_{\text{g}}^{\text{HSE}}- \text{E}_{\text{g}}^{\text{PBE+U}})^2-{\alpha_2}(\Delta \text{Band})^2
\end{equation}
Here, $\vec{U}$ = [$U^{1}$, $U^{2}$,...,$U^{n}$] is the vector of $U_{eff}$ values applied to different atomic species and $U^{i} \in [-10, 10]$ eV. $\Delta \text{Band}$ is defined as the mean squared error of the PBE+U band structure with respect to HSE:
\begin{equation}
{\Delta \text{Band}}=\sqrt{\frac{1}{N_E}\sum^{N_k}_{i=1}\sum^{N_b}_{j=1}(\epsilon_{HSE}^{j}[k_i]-\epsilon_{PBE+U}^{j}[k_i])^2}
\end{equation}
$N_E$ represents the total number of eigenvalues,  $\epsilon$, included in the comparison, $N_k$ is the number of $k$-points, and $N_b$ is the number of bands selected for comparison. To avoid double counting the band gap difference in the calculation of $\Delta \text{Band}$, the valence band maximum (VBM) and conduction band minimum (CBM) are shifted to zero for both the PBE+U and HSE band structures. Hence, $\Delta \text{Band}$ captures differences in the qualitative features of the band structures produced by PBE+U vs. HSE, independently of the difference in the band gap.
The coefficients $\alpha_1$ and $\alpha_2$ may be used to assign different weights to the band gap vs. the band structure. The default values are 0.25 and 0.75, respectively. For CdTe, we set $\alpha_1 = \alpha_2 = 0.5$ to assign a higher weight to the band gap term. For HgTe, which is a metal, we set $\alpha_1 = 0$ and $\alpha_2 = 1$.

\begin{figure}[H]
\includegraphics[scale = 0.0785]{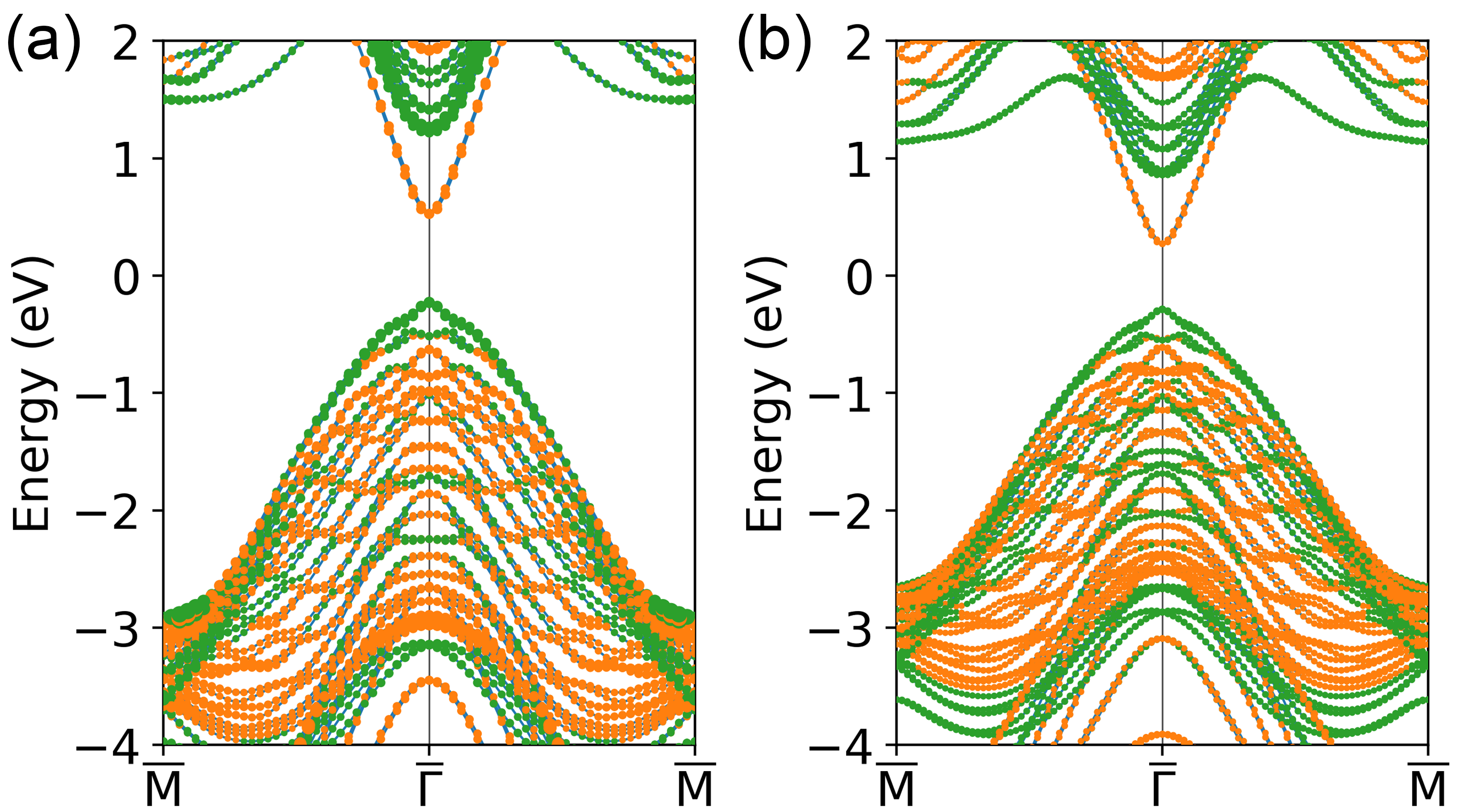}
\caption{\label{fig:hse} The band structure of an InAs/GaSb interface with 5 layers of InAs and 5 layers of GaSb obtained with (a) HSE and (b) DFT+U(BO). Orange and green dots indicate the contributions of InAs and GaSb, respectively.}
\end{figure}

For InAs and GaSb the optimal values of  U$_{eff}$ have been found to be:  U$_{eff}^{In, p}$ = -0.5 eV, U$_{eff}^{As, p}$ = -7.5 eV,  U$_{eff}^{Ga, p}$ = 0.8 eV, U$_{eff}^{Sb, p}$ = -6.9 eV, as reported in \cite{yu2020machine}. With these parameters, DFT+U(BO) yields a band gap of 0.31 eV for InAs, in good agreement with the experimental value of 0.41 eV \cite{vurgaftman2001band}, and a band gap of 0.45 eV for GaSb, which is somewhat underestimated compared to the experimental value of  0.81 eV \cite{vurgaftman2001band}.
For CdTe, BO produces an optimal value of U$_{eff}^{Cd, d}$ = 8.3 eV, somewhat higher than the value of 7 eV used in \cite{wu2015lda+}. This results in a band gap of 0.87 eV, which is closer to experiment than previous \textit{ab initio} calculations \cite{heyd2005energy,wei2000first}. The qualitative features of the PBE+U(BO) band structure are in agreement with HSE, as shown in Fig. \ref{fig:hgte_hse}e, however the gap and the band width are still somewhat underestimated. For HgTe, BO produces a value of U$_{eff}^{Hg, d}$ = 8.4 eV, somewhat lower than the value of 9.4 eV used in Ref. \cite{PhysRevB.90.195311}. The band structure, shown in Fig. \ref{fig:hgte_hse}d,  has the correct band shape, comparable to the HSE band structure, and is in agreement with Ref. \cite{PhysRevB.90.195311}.  To demonstrate the transferability of the U$_eff$ values obtained by BO from bulk materials to interfaces, we compare the band structures produced by PBE+(BO) and HSE for an InAs/GaSb interface with 5 layers of InAs and 5 layers of GaSb, constructed as detailed below. Fig \ref{fig:hse} shows that overall good agreement is obtained between PBE+U(BO) and HSE, however PBE+U(BO) somewhat underestimates the band gap and the band width. We note that the U$_{eff}$ values obtained here are based on the implementation of the Dudarev formalism in VASP. Different DFT+U implementations may yield different results \cite{Harald_DFT+U,jiang2010first}. 

\subsection{Interface model construction}

For the HgTe/CdTe(100) interface, we constructed periodic heterostructures, similar to Ref. \cite{PhysRevB.90.195311}. However, we used a larger number of  CdTe layers to ensure convergence, as detailed below.  The thickness of HgTe was varied to study the evolution of the electronic structure. The experimental lattice constants of 6.45 \si{\angstrom} for HgTe and 6.48 \si{\angstrom} for CdTe are closely matched \cite{west1999basic}. We assumed that an epitaxially matched HgTe film  would grow on top of a CdTe substrate with the experimental lattice constant of 6.48 \si{\angstrom}.  

For the InAs/GaSb interface, we constructed two types of interface slab models: The InSb-type interface has In and Sb as the terminal atoms at the surfaces and interface.  The GaAs-type interface has Ga and As as the terminal atoms. The experimental lattice constants of 6.058 \si{\angstrom} for InAs and 6.096 \si{\angstrom} for GaSb \cite{vurgaftman2001band} are closely matched.  We assumed that an epitaxially matched InAs film would grow on top of GaSb with the lattice constant of 6.096 \si{\angstrom}, based on the experiment in Ref. \cite{shojaei2018materials}. To study the effect of the InAs and GaSb thickness, interface models were constructed with the number of layers of each material varying from 10  to 50. The notation "A/B" is used to describe an InAs/GaSb interface with A layers of InAs and B layers of GaSb. A vacuum region of about 40 \si{\angstrom} was added to the interface model to prevent spurious interactions between periodic images (for the purpose of band unfolding the closest integer number of primitive cells to 40 \si{\angstrom} was used \cite{yang2020electronic}). In order to terminate dangling bonds, In and Ga atoms on the surface were passivated by pseudo hydrogen atoms with 1.25 fractional electrons, whereas As and Sb atoms on the surface were passivated by pseudo hydrogen  atoms with 0.75 fractional electrons. Structural relaxation was performed for the surface atoms and passivating pseudo-hydrogen atoms until the change of the all forces was below 10$^{-3}$ eV/\si{\angstrom}. 

\begin{figure}
\includegraphics[scale = 0.5]{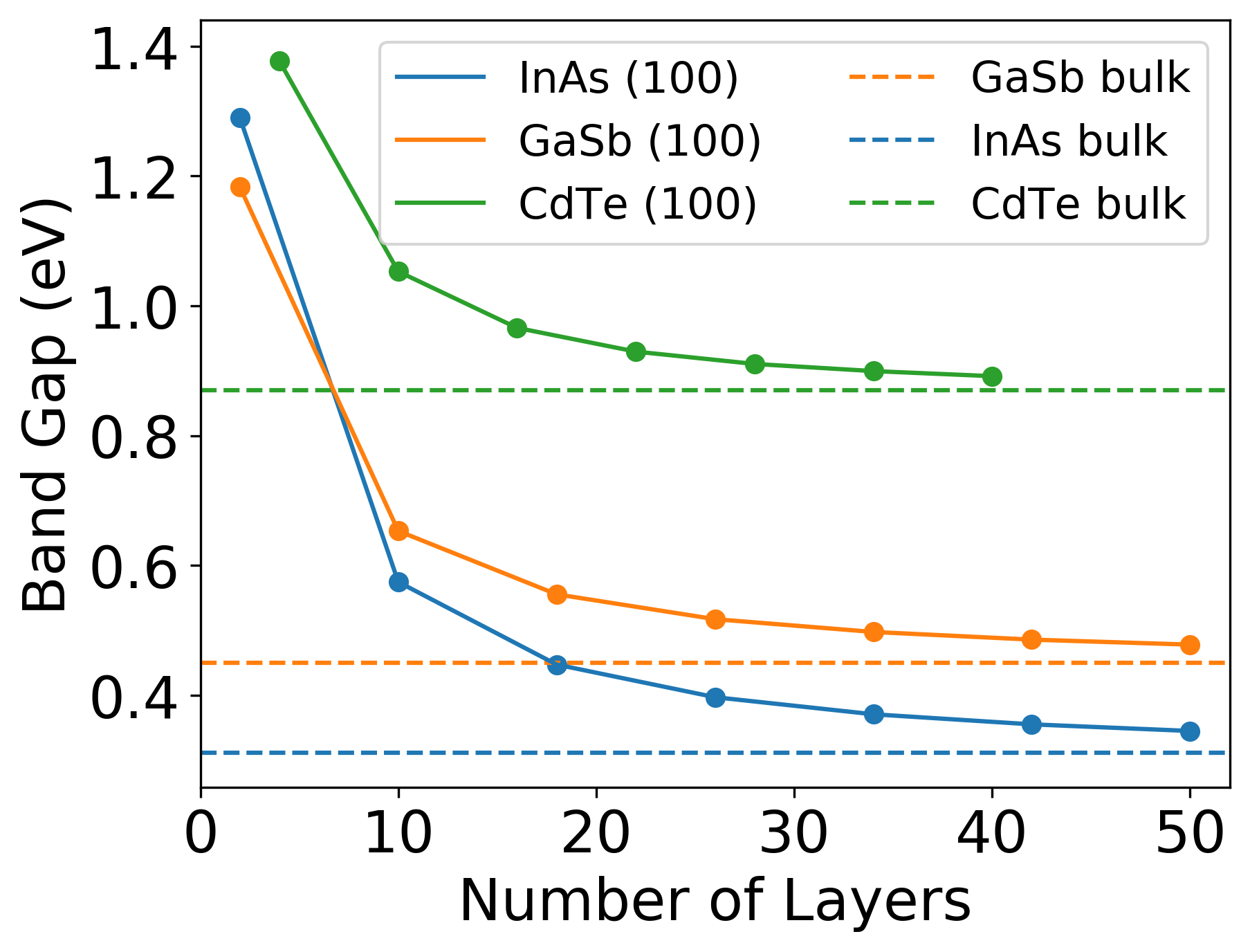}
\caption{\label{fig:convergence} The band gap obtained with PBE+U(BO) as a function of the number of layers for InAs(100), GaSb(100), and CdTe(100) surface slabs. }
\end{figure}

The number of layers included in slab models needs to be converged to the bulk limit to avoid quantum size effects. For semiconductors the band gap is typically used as a the convergence criterion \cite{PhysRevMaterials.4.034203, yang2020electronic}.
Fig. \ref{fig:convergence}  shows  the  band  gap  as  a  function  of  the number  of  layers  for  InAs(100), GaSb(100), and CdTe(100).  We note that here "layer" is defined as one atomic layer.  In each iteration, the number of layers was increased by 8 for InAs and GaSb and by 6 for CdTe.  If the band gap difference between the current iteration and the previous iteration was within $1\times10^{-2}$ eV, the current number of layers was regarded as converged.  For InAs and GaSb surfaces, 50 layers are required,  whereas for CdTe 40 layers are required to converge the band gap. The converged band gap values are close to the bulk values. The size of the interface models used to simulate the effect of an electric field was limited to 10 layers of InAs with 10 layers of GaSb due to convergence issues, as detailed in the SI.

\section{RESULTS AND DISCUSSION}
\subsection{HgTe/CdTe }

\begin{figure*}
\includegraphics[scale =  0.075]{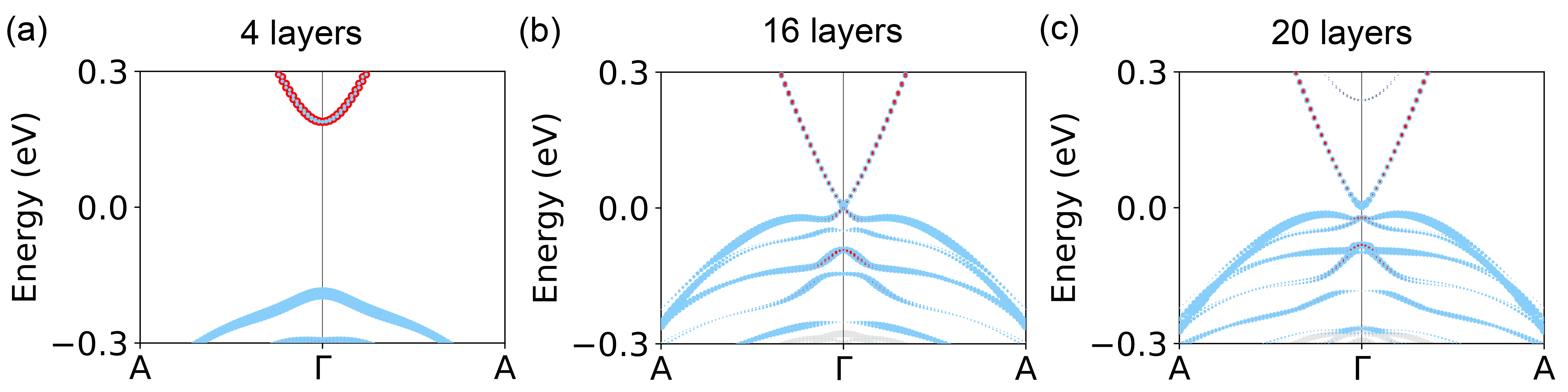}
\caption{\label{fig:hgte_bands} Band structures of a HgTe/CdTe interface with 40 layers of CdTe and (a) 4 layers, (b) 16 layers, and (c) 20 layers of HgTe. The red dots indicate the contributions of Hg $s$ states and the blue dots indicate the contributions of the Te $p$ states.}
\end{figure*}

\begin{figure}
\includegraphics[scale = 0.5]{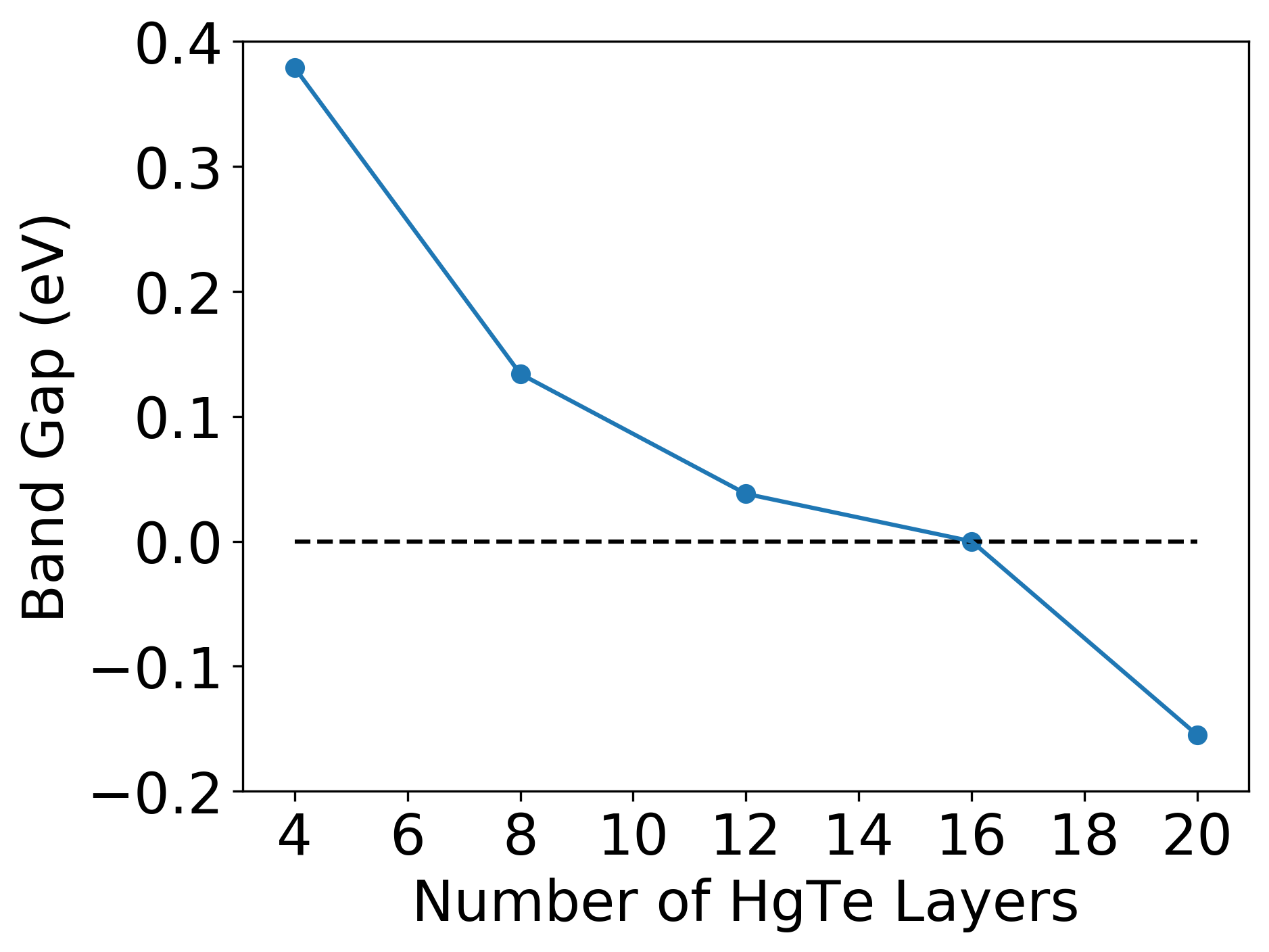}
\caption{\label{fig:hgte_gap} The band gap of a HgTe/CdTe interface with 40 layers of CdTe as a function of the number of HgTe layers. Negative values indicate an inverted band gap. }
\end{figure}

To validate the DFT+U(BO) method, we begin by applying it to the well-studied HgTe/CdTe interface. Bulk-unfolded band structures of HgTe/CdTe heterostructures with 40 layers of CdTe and a varying number of HgTe layers are shown in Fig. \ref{fig:hgte_bands}. The red dots indicate the contributions from Hg $s$ orbitals and the blue dots indicate the contributions from Te $p$ orbitals. The band gap value as a function of the number of HgTe layers is shown in Fig. \ref{fig:hgte_gap}. Negative values indicate an inverted band gap. A drastic change is observed with the thickness of HgTe. When the number of layers is below 16, the interface behaves as a trivial insulator, with the Hg $s$ orbitals forming the bottom of the conduction band and the Te $p$ orbitals forming the top of the valence band. When the number of HgTe layers reaches 16, a transition point from a trivial insulator to a topological insulator occurs. At this transition point, both the CBM and VBM show a hybridized $sp$ character. When the number of HgTe layers exceeds 16, an inverted gap opens, leading to the occurrence of a topologically nontrivial phase, in which the VBM is dominated by Hg $s$ states and the CBM is dominated by Te $p$ states. The critical thickness of 16 layers, corresponds to 5.1 nm in good agreement with the experimental result of 6.3 nm (around 19 layers) \cite{bernevig2006quantum}. Our result is comparable to previous DFT calculations, which used different functionals and considered structures with fewer layers of CdTe. Ref. \cite{PhysRevB.89.195312} obtained a critical thickness of 4.6 nm of HgTe on top of 4 layers of CdTe using the modified Becke-Johnson (MBJ) functional.Ref. \cite{PhysRevB.90.195311} obtained a critical thickness of 6.5 nm of HgTe on top of 10 layers of CdTe using GGA+U for HgTe and GGA for CdTe. Thus, the DFT+U(BO) method successfully describes the electronic structure of the HgTe/CdTe interface and captures the transition from trivial to topological behavior.

\subsection{InAs/GaSb}
\subsubsection{Effect  of  layer  thickness}
\begin{figure}[H]
\includegraphics[scale = 0.55]{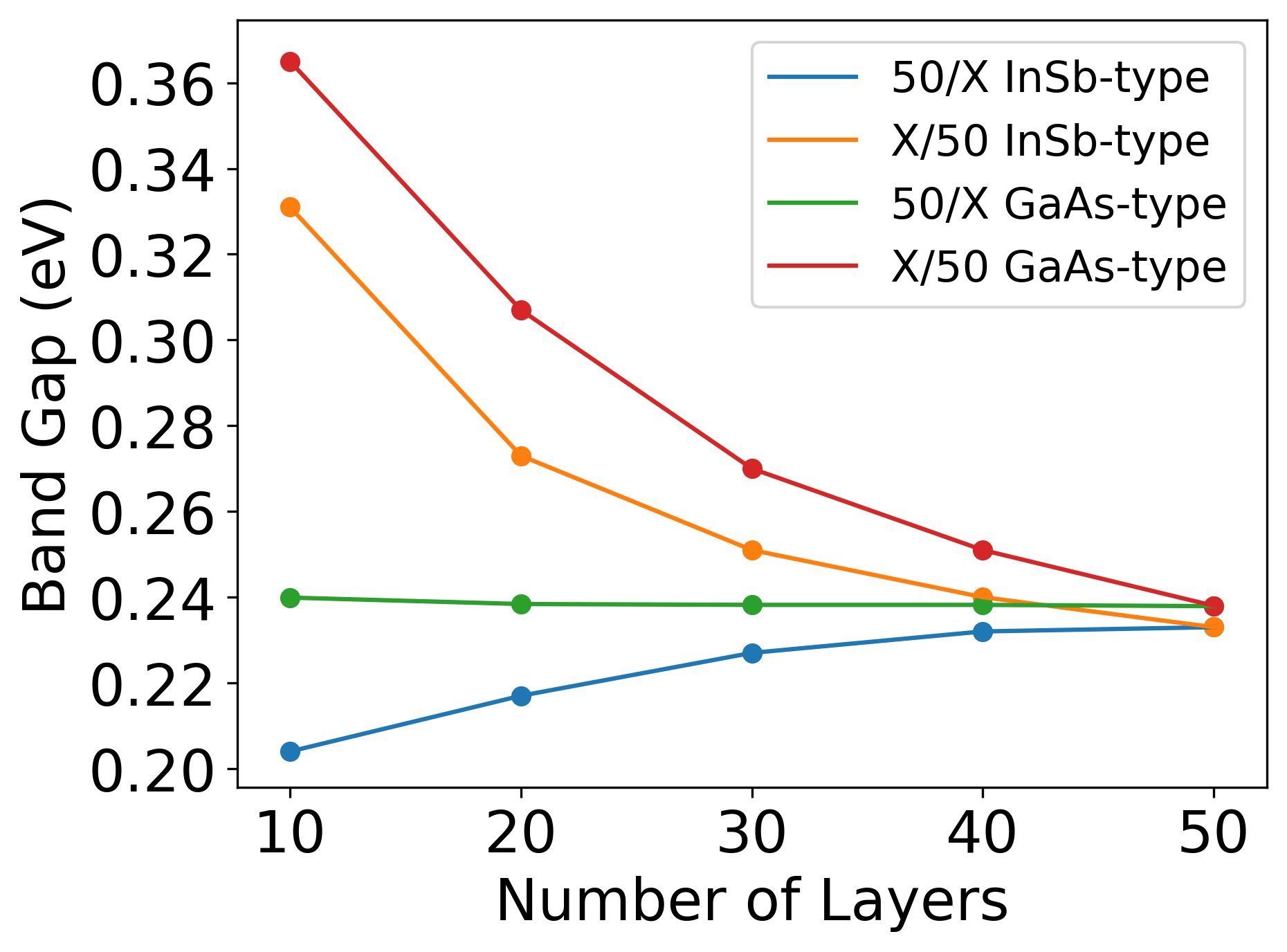}
\caption{\label{fig:gap} Band gap values as a function of number of layers for 50-layer InAs/X-layer GaSb and X-layer InAs/50-layer GaSb of InSb-type and GaAs-type interface. }
\end{figure}
To investigate the influence of the thickness of InAs and GaSb on the band gap, we conducted two series of calculations for InSb-type and GaAs-type interfaces. In one series, the thickness of InAs was fixed at 50 layers InAs and the number of GaSb layers (X) was varied. In the other series, the thickness of GaSb was fixed at 50 layers and the number of InAs layers (X) was varied. The results are shown in Fig. \ref{fig:gap}. For the InAs(50)/GaSb(X) series, the band gap of the InSb-type interface increases with increasing GaSb thickness, whereas the band gap of the GaAs-type interface does not change significantly. For the InAs(X)/GaSb(50) series, the band gap decreases with increasing InAs thickness for both interface types, although the gap of the InSb-type interface remains smaller than that of the GaAs-type interface throughout. The trend of the gap decreasing with the increase in InAs thickness is in agreement with experimental observations \cite{klein2012temperature}. The thickest interface we were able to calculate comprises 50 layers, which corresponds to about 15 nm of each material. The bulk-unfolded band structure of a 50/50 InSb-type interface is shown in the SI. 
 Because this interface still has a gap of over 0.2 eV, and the rate of the gap narrowing decreases with increasing InAs thickness, as shown in Fig. \ref{fig:gap}, we estimate that it would either require a significantly thicker film of InAs for the gap to completely close or the gap would approach a finite asymptotic limit rather than close. In addition to increasing the QW thickness, strain engineering, which is not taken into account here, may also help modulate the gap.~\cite{akiho2016engineering, du2017tuning, tiemann2017impact}.

We note that an analysis based on the empirical 8-band Kane model found band inversion and the quantum spin Hall phase for an InAs thickness above 9\,nm at fixed 10\,nm GaSb thickness~\cite{liu2008quantum}. However, this analysis was based on empirical parameters for the material and interface properties and did not take the atomic details of the interface structure into account. For example, in Ref.~\cite{liu2008quantum} the band alignment at the interface was chosen such that the GaSb valence band is 150\,meV higher than the InAs conduction band leading to a band inversion even for relatively thin layers. In contrast, within our first principles approach, we find that band inversion is not achieved up to an InAs thickness of 15\,nm for a range of GaSb thicknesses including 10\,nm. Furthermore, we found that the atomic details of the interface, like the type of bonds formed at the interface (InSb or GaAs), are relevant, which was neglected in the effective theory of Ref.~\cite{liu2008quantum}. Finally, it should be noted that experiments seem to indicate that an electric field is required to achieve an inverted regime in InAs/GaSb heterostructures~\cite{qu2015electric}.

\subsubsection{Effect  of  electric field}
\begin{figure*}
\includegraphics[scale = 0.075]{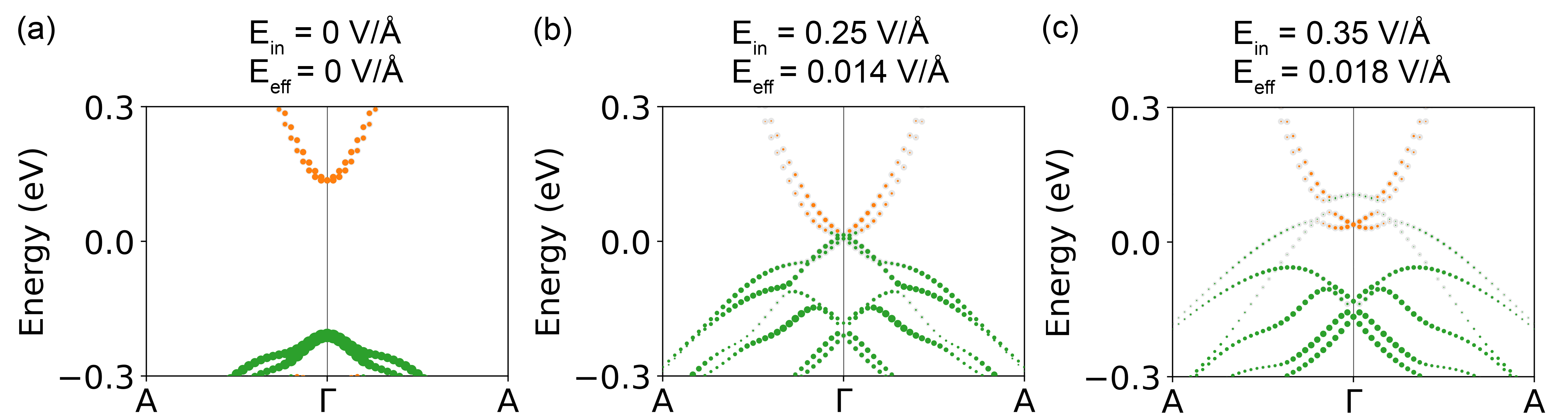}
\caption{\label{fig:electric_insb} Electronic structure of a 10/10 InSb-type interface with different external electric fields. a-c) Bulk unfolded band structures with the contributions of the interface layers of InAs and GaSb colored in orange and green, respectively.}
\end{figure*}

The band alignment at the interface of InAs/GaSb can be manipulated by applying external gate voltages. Ref.\cite{qu2015electric} has presented strong experimental evidence that the gap closes when the external gate voltages reach a critical value. Therefore, we performed DFT simulations for interface slabs in presence of electric field. In the VASP code, an external electric field is simulated by adding an artificial dipole sheet in the vacuum region of the unit cell \cite{PhysRevB.46.16067}. Due to screening effects and the electric susceptibility inside the materials, the effective electric field at the interface may be significantly smaller than the input electric field \cite{PhysRevB.46.16067,wang2018electric}. To estimate the effective electric field, we calculated the gradient of the potential in the InAs and in GaSb, based on the electrostatic potential averaged over xy plane. The averaged gradient is taken as the effective electric field. The full account of the effective field estimation is provided in the SI. The electric field is applied perpendicular to the plane of the interface and points from the GaSb side to the InAs side. We note that in VASP only an external electric field can be set, whereas in experiments the position of the Fermi level can be independently controlled by applying front-gate and back-gate voltages. Owing to convergence issues in DFT calculations with external electric fields (see SI), the largest interfaces we were able to calculate comprise 10 layers of InAs and 10 layers of GaSb.  

\begin{figure}
\includegraphics[scale = 0.5]{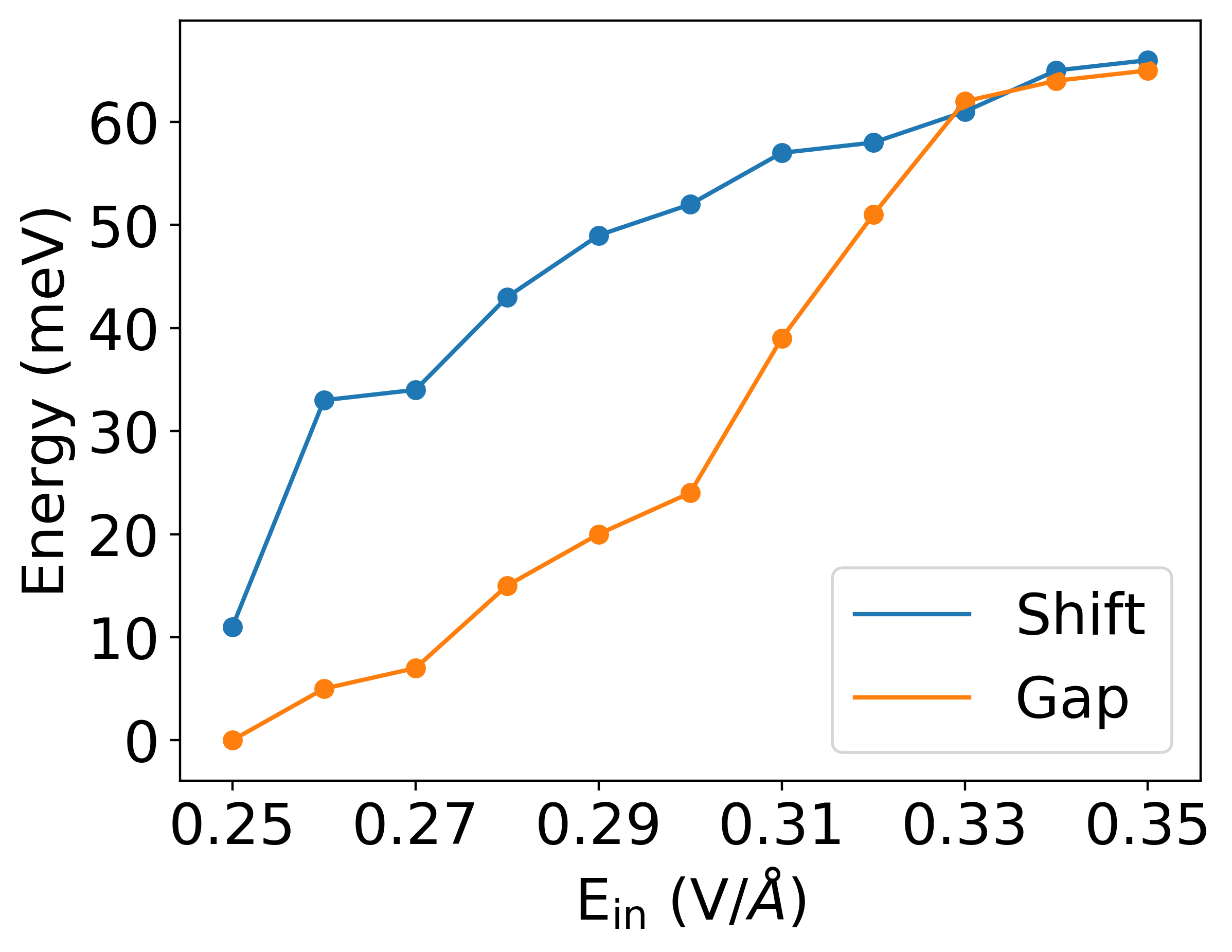}
\caption{\label{fig:distance} The inverted band gap at $\Gamma$ and its position above the Fermi level as a function of the input electric field,  where the shift is defined as the energy difference between the position of the bottom of the inverted band gap and the Fermi level at the $\Gamma$ point.}
\end{figure}

Fig. \ref{fig:electric_insb} shows the band structure of a 10/10 InSb-type interface. When no electric field is applied, the interface is in the trivial insulator state. The CBM is dominated by the interface InAs layer (orange), whereas the interface GaSb layer (green) contributes predominantly to the VBM. As the electric field increases, the bands contributed by the GaSb shift upwards with respect to the bands contributed by the InAs and the gap narrows. When the input electric field reaches 0.25 V/\si{\angstrom}, which corresponds to an effective field of 0.014 V/\si{\angstrom}, the GaSb VBM overlaps with the InAs CBM, the gap closes, and band crossing occurs. Our results are qualitatively in agreement with previous studies \cite{naveh1995band,liu2008quantum, qu2015electric}, which indicated that the band gap in InAs/GaSb could be closed via an external electric field. When the electric field is increased further, an inverted gap opens. As the electric field is increased, the inverted gap expands, but also shifts higher above the Fermi level. Fig \ref{fig:distance} shows the position of the inverted gap above the Fermi level at the $\Gamma$ point as a function of the electric field. With an input electric field of 0.35 V/\si{\angstrom}, which corresponds to an effective field of 0.018 V/\si{\angstrom}, the gap at the $\Gamma$ point is 65 meV and the bottom of the inverted gap is found 66 meV above the Fermi level.  For the GaAs-type interface, shown in the SI, the band gap also decreases as the electric field increases. However, because the GaAs-type interface has a larger band gap and the effect of the electric field is weaker than for the InSb-type interface, the gap does not close even for an input electric field as high as 0.55~V/\si{\angstrom}.

\begin{figure}
\includegraphics[scale = 0.5]{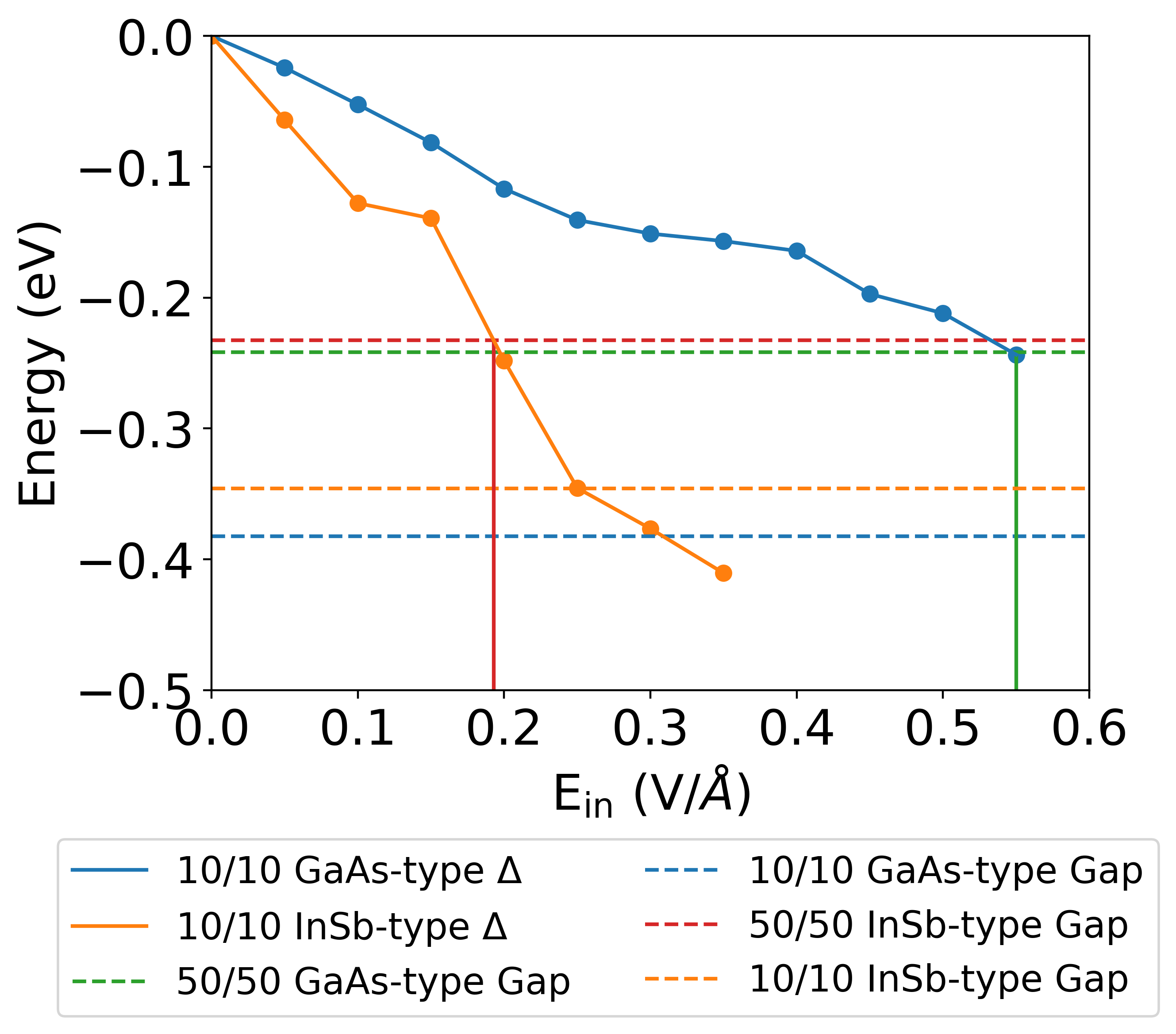}
\caption{\label{fig:trend} The band gap reduction, $\Delta$, as a function of electric field for 10/10 GaAs-type and InSb-type interfaces.  }
\end{figure}

Fig \ref{fig:trend} shows the change in the band gap, $\Delta$, as a function of the input electric field, $E_{in}$, for 10/10 GaAs-type and InSb-type interfaces:
\begin{equation}
    \Delta = Gap(E_{in}) - Gap(E_{in}=0) 
\end{equation}
The blue and orange dashed lines indicate the band gaps of the 10/10 GaAs-type and InSb-type interfaces, respectively. The gap closes when the dashed line is crossed. To estimate the input electric field that would be required for the gap to close for a 50/50 interface, we assume that the change in the gap would behave similarly to a 10/10 interface. The green and red dashed lines indicate the band gaps of the 50/50 GaAs-type and InSb-type interfaces, respectively. Based on this, we estimate that an input electric field of 0.19 V/\si{\angstrom}, which corresponds to an effective electric field of 0.012 V/\si{\angstrom}, would be needed to close the gap for a 50/50 InSb-type interface, as indicated by the red solid line. For the GaAs-type interface an input electric field of 0.55 V/\si{\angstrom}, which corresponds to an effective electric field of 0.017 V/\si{\angstrom}, would be needed to close the gap, as indicated by the green solid line.  We highlight that the effective electric field of 0.017~V/\si{\angstrom} corresponds to a potential drop of $2.55$~V over the 15 nm thickness of the QW in this case, which is likely to make the material conducting well before the topological transition. Our results indicate that while it may be possible to tune the InAs/GaSb interface into the topological regime, it would not be trivial.

\section{Conclusion}
In summary, we have studied the HgTe/CdTe and InAs/GaSb quantum wells using DFT with a Hubbard U correction determined by Bayesian optimization. DFT+U(BO) produces band structures of comparable accuracy to a hybrid functional at the computational cost of a semi-local functional. This enables us to conduct simulations of large interface models with hundreds of atoms. 

For the HgTe/CdTe interface we find that an inverted gap opens at a critical thickness of 5.1 nm of HgTe, in agreement with experimental observations and previous theoretical studies. For InAs/GaSb QWs with 50 layers (about 15 nm) of GaSb we find that the gap narrows with increasing thickness of InAs in agreement with the previous theory estimations.
However, the gap does not completely close with up to 50 layers (about 15 nm) of InAs. Based on the rate of gap narrowing, we estimate that it would either require a significantly thicker InAs film to close the gap or the gap would decay to a finite asymptotic limit. 

Simulations with an external electric field applied perpendicular to the interface, pointing from GaSb to InAs, have been conducted for models with 10 layers of each material. We find that with increasing field strength the GaSb VBM shifts upwards relative to the InAs CBM, leading to narrowing of the gap at the interface. For the InSb-type interface, band crossing is observed at a critical field and subsequently an inverted gap opens. As the electric field increases the gap increases but also shifts higher in energy above the Fermi level. Because the 10/10 interface has a larger gap due to the quantum size effect, we estimate the reduced critical field that would be required to achieve band inversion and reach the topological regime for thicker QWs comprising 50 layers of each material.  

Our results explain the difficulty of experimentally reaching the topological regime in InAs/GaSb QWs. In principle, under the right conditions, an inverted gap could be produced in this system. However, achieving this requires a delicate balance between several parameters. To tune the initial gap, the structure of the QWs must be precisely controlled, including the layer thickness, the bonding configuration at the interface, and possibly also the lattice strain. Even if a smaller zero-field gap is obtained by interface engineering, a considerable electric field may still be required to obtain band crossing and drive the system into the topological regime. Finally, gating or doping may be required to tune the Fermi level position inside the inverted gap.    


The HgTe/CdTe QW does not suffer from this difficulty because HgTe has an inverted band structure intrinsically. Therefore, no electric field is necessary to achieve band inversion at the HgTe/CdTe interface and it is easier to reach the topological regime.
Thus, our results make a case for limited applicability of InAs/GaSb quantum wells for 2DTI production and suggest that alternative, more promising materials should be sought.



\begin{acknowledgments}
We would like to thank Sergey Frolov from the University of Pittsburgh, Chris Palmstr{\o}m from the University of California, Santa Barbara, Vlad Pribiag from the University of Minnesota, and Michael Wimmer from TU Delft for helpful discussions. Work at CMU was funded by the National Science Foundation (NSF) through grant OISE-1743717. This research used resources of the National Energy Research Scientific Computing Center (NERSC), a DOE Office of Science User Facility supported by the Office of Science of the U.S. Department of Energy under contract no. DE-AC02-05CH11231.
\end{acknowledgments}

\end{document}